\documentclass[prb,floatfix,showpacs,tightenlines,groupedaddress,superscriptaddress]{revtex4}
\usepackage{amsmath}
\usepackage{graphicx}
\usepackage{epsfig}

\begin{document}

\title{Charge transport in a quantum electromechanical system}

\author{D. Wahyu Utami}
\affiliation{Center for Quantum Computer Technology and Department
of Physics, School of Physical Sciences, The University of
Queensland, QLD 4072, Australia}
\author{Hsi-Sheng Goan}
\affiliation{Center for Quantum Computer Technology,
University of New South Wales, Sydney, NSW 2052,
Australia}
\author{G.~J.~Milburn}
\affiliation{Center for
Quantum Computer Technology and Department of Physics, School of
Physical Sciences, The University of Queensland, QLD 4072,
Australia}

\date{\today}

\begin{abstract}
We describe a quantum electromechanical system(QEMS)
comprising a single quantum dot harmonically bound between two
electrodes and facilitating a tunneling current between them. An
example of such a system is a fullerene molecule between two metal
electrodes [Park et al. Nature, {\bf 407}, 57 (2000)].  The description
is based on a quantum master equation for the density operator of the
electronic and vibrational degrees of freedom and thus incorporates
the dynamics of both diagonal (population) and off diagonal
(coherence) terms. We derive coupled equations of motion for the
electron occupation number of the dot and the vibrational degrees of
freedom, including damping of the vibration and thermo-mechanical
noise.  This dynamical description is related to
observable features of the system including the stationary
current as a function of bias voltage.
\end{abstract}

\maketitle

\section{Introduction}
A quantum electromechanical system (QEMS) is a submicron
electromechanical device fabricated through state-of-the-art
nanofabrication\cite{roukes}. Typically, such
devices comprise a mechanical oscillator (a singly or doubly clamped
cantilever) with surface wires patterned through shadow mask metal
evaporation. The wires can be used to drive the mechanical system by
carrying an AC current in an external static magnetic field. Surface
wires can also be used as motion transducers through induced EMFs as
the substrate oscillates in the external magnetic field.
Alternatively the mechanical resonators can form an active part of a
single electron transducer, such as one plate of a capacitively
coupled single electron transistor\cite{knobel-clleland,Schwab04}. These
devices have been proposed as sensitive weak force probes with the
potential to achieve single spin
detection\cite{spin-detection,Brun03}.
However they are of considerable
interest in their own right as nano-fabricated mechanical resonators
capable of exhibiting  quantum noise features, such as squeezing and
entanglement \cite{quantum-effects}.

In order to observe quantum noise in a QEMS device we must
recognize that these devices are open quantum systems and
explicitly describe the interactions between the device and a
number of thermal reservoirs. This is the primary objective of
this paper. There are several factors that determine whether a
system operates in the quantum or classical regime. When the
system consists only of an oscillator coupled to a bath the
oscillator quantum of energy should be greater than the
thermo-mechanical excitation of the system; $\hbar\omega_o\geq
k_BT$ where $\omega_o$ is the resonant frequency of the QEMS
oscillator and $T$ is the temperature of the thermal mechanical
bath in equilibrium with the oscillator. At a temperature of 10
milliKelvin, this implies an oscillator frequency of the order of
GHz or greater. Recently Huang et al. reported the operation of a
GHz mechanical oscillator\cite{Huang}. A very different approach
to achieving a high mechanical frequency was the fullerene
molecular system of Park et al.\cite{Park}, and it is this system
which we take as the prototype for our theoretical description.
Previous work on the micro-mechanical
degrees of freedom coupled to mesoscopic conductors
\cite{Mozyrsky-Martin, Smirnov, Mozyrsky-Martin-Hastings,Armour},
indicate that
transport of carriers between source and drain can act as a damping
reservoir, even in the absence of an other explicit mechanism for
mechanical damping into a thermal reservoir. This is also predicted by
the theory we present for a particular bias condition.
As is well known, dissipation can  restore semiclassical
behavior.  Transport induced damping can also achieve this result.

\begin{figure}
\centerline{\includegraphics[width=6cm]{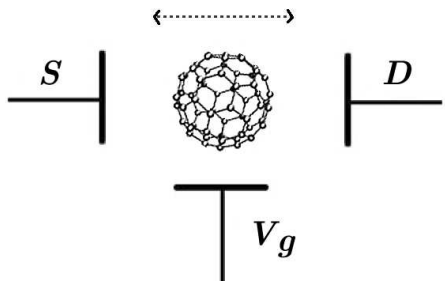}}
\caption{Schematic representation of tunneling between a source and a
  drain through a quantum dot. The dot is harmonically bound and
  vibrational motion can be excited as electrons tunnel through the
  system.}
\label{fig-1}
\end{figure}

The model we describe, Fig.~\ref{fig-1}, consists of a single
quantum dot coupled via tunnel junctions to two reservoirs, the
source and the drain. We will assume that the  coulomb blockade
permits only one quasi-bound single electron state on the dot
which participates in the tunneling between the source and the
drain. We will ignore spin, as the source and drain are not spin
polarized, and there is no external magnetic field.  A gate
voltage controls the energy of this quasi-bound state with respect
to the Fermi energy in the source.   The quantum dot can oscillate
around an equilibrium position mid way between the source and the
drain contacts due to weak restoring forces. When an electron
tunnels  onto the dot an electrostatic force is exerted on the dot
shifting its equilibrium position.  In essence this is a quantum
dot single electron transistor.  In the experiment of Park et
al.\cite{Park}, the quantum dot was a single fullerene molecule
weakly bound by van der Walls interactions between the molecule
and the electrodes. The dependence of the conductance on gate
voltage was found to exhibit features attributed to transitions
between the quantized vibrational levels of the mechanical
oscillations of the molecule.

Boese and Schoeller\cite{Boese} have recently given a theoretical
description of the conductance features of this system. A more detailed analysis using similar techniques was given by Aji et al.\cite{Aji}.  Our objective
is to extend these models to provide a full master equation description
of the irreversible dynamics,  including quantum
correlation between the mechanical and  electronic degrees of
freedom.  We wish to go beyond a rate equation description so as to be
able to include coherent quantum effects which arise, for example, when the mechanical degree of freedom is subject to coherent driving.

\section{The model}
We will assume that the center of mass of the dot is  bound in a
harmonic potential with resonant frequency $\omega_o$.  This
vibrational degree of freedom is then described by a displacement
operator $\hat{x}$ which can be written in terms of annihilation and
creation operators $a,a^\dagger$ as
\begin{equation}
\hat{x}=\sqrt{\frac{\hbar}{2m\omega_o}}(a+a^\dagger).
\end{equation}
The electronic single quasi-bound state on the dot is
described by Fermi  annihilation and creation operators $c,c^\dagger$,
which satisfy the anti commutation relation $cc^\dagger +c^\dagger
c=1$.

The Hamiltonian of the system can then be written as,
\begin{eqnarray}
\lefteqn{H= \hbar \omega_I (V_g) c^{\dagger} c + U_c \hat{n}^2}
\\
& &+ \hspace{0.1cm} \hbar \omega_o a^{\dagger} a \label{e:osc}\\
& &+ \hspace{0.1cm} \hbar\sum_k \omega_{Sk} a_k^{\dagger} a_{k} + \hbar
\sum_k\omega_{Dk} b_k^{\dagger}
b_k \label{e:dns}\\
& &- \chi ( a^{\dagger} + a) \hat{n} \label{e:coupling}\\
& &+ \sum_k T_{Sk} (a_k c^\dagger + c a_k^\dagger) + \sum_k T_{Dk} (b_k c^\dagger + c b_k^\dagger) \label{e:dnscoupling}\\
& &+ \sum_p \hbar \omega_p d_p^\dagger d_p + g_p (a^\dagger d_p + a
d_p^\dagger) \label{e:damp},
\end{eqnarray}
where $\hat{n}=c^\dagger c$ is the excess electron number operator on the dot.

The first term of the Hamiltonian describes the free energy for
the island. A particular gate voltage $V_g$, with a corresponding
$\hbar \omega_I = 15 meV$, for the island is chosen for
calculation. $U_c$ is the Coulomb charge energy which is the
energy that is required to add an electron when
there is already one electron occupying the island. We will assume
this energy is large enough so that no more than one electron
occupies the island at any time. This is a Coulomb Blockade
effect. The charging energy of the fullerene molecule transistor
has been observed by Park et al.\ to be larger than 270 meV which is two
orders of magnitude larger than the vibrational quantum of energy
$\hbar \omega_o$. The free Hamiltonian for the oscillator is
described in term (\ref{e:osc}). The Park et al.\ experiment gives
the value $ \hbar \omega_o= 5$ meV, corresponding to a THz oscillator.
The electrostatic energy of electrons in the source and drain reservoirs is
 written as term (\ref{e:dns}). Term (\ref{e:coupling}) is the
coupling between the oscillator and charge while term
(\ref{e:dnscoupling}) represents the source-island tunnel
coupling and the drain-island tunnel coupling. The last term,
(\ref{e:damp}), describes the coupling between the oscillator and
the thermo-mechanical bath responsible for damping and thermal
noise in the mechanical system in the rotating wave
approximation\cite{Gardiner-Zoller}. This is an additional source of
damping to that which can arise due to the transport process itself
(see below). We include it in order to bound the motion under certain
bias conditions. A possible physical origin of this source of
dissipation will be discussed after the derivation of the master
equation.

We have neglected the position dependence of the tunneling rate
onto and off the island. This is equivalent to assuming that the
distance, $d$, between the electrodes and the equilibrium position
of the uncharged quantum dot, is much larger than the rms position
fluctuations in the ground state of the oscillator.  There are
important situations where this approximation cannot be made, for
example in the so called `charge shuttle' systems
\cite{shuttle-rev}.

A primary difficulty in analyzing the quantum dynamics of this open
system is the presence of different time scales associated with the
oscillator, the tunneling events and the coupling between the
oscillator and electronic degrees of degree due to the electrostatic
potential, term (4). The standard approach
would be to move to an interaction picture for the oscillator and the
electronic degrees of freedom. However this would make the
electrostatic coupling energy time dependent, and rapidly oscillating.
Were we to approximate this with the secular terms stemming from a
Dyson expansion of the Hamiltonian, the resulting effective coupling
between the oscillator and the electron occupation of the dot simply
shifts the free energy of the dot and no excitation of the mechanical
motion can occur.

To avoid this problem we eliminate the
coupling term of the oscillator and charge by doing a canonical
transformation with unitary representation  $U=e^s$ where,
\begin{equation}
s = -\lambda (a^\dagger -a) \hat{n}
\end{equation}
with
\begin{equation}
  \label{eq:lambda}
  \lambda=\frac{\chi}{\hbar\omega_o}\; .
\end{equation}
This unitary transformation gives a conditional displacement of the
oscillator conditional on the electronic occupation the dot. One might
call this a {\em displacement} picture.

This derivation follows the approach of Mahan \cite{mahan}. The
motivation behind this is as follows. The electrostatic interaction,
term (4), displaces the equilibrium position of the oscillator so that
the average value of
the oscillator amplitude in the ground state becomes
\begin{equation}
\langle a\rangle=\lambda,
\end{equation}
We can shift this back to the origin by a phase-space displacement
\begin{equation}
\bar{a}\equiv e^{s}ae^{-s}=a+\lambda \hat{n}.
\end{equation}
This unitary transformation gives a conditional displacement of the
oscillator, conditional on the electronic occupation of the
dot. Applying $U$ to the Fermi operator $c$ gives
\begin{equation}
\bar{c} = c e^{\lambda (a^\dagger -a)}.
\end{equation}

The Schr\"{o}dinger equation for the displaced state,
$\bar{\rho}=e^s\rho e^{-s}$, then takes the form
\begin{equation}
\frac{d\bar{\rho}}{dt}=-\frac{i}{\hbar}[\bar{H},\bar{\rho}],
\end{equation}
where  the transformed Hamiltonian is
\begin{eqnarray}
\bar{H}=\hbar \omega_o a^{\dagger} a + \hbar \omega_I (V_g) c^{\dagger} c +\sum_k\hbar \omega_{Sk} a_k^{\dagger} a_k +
\sum_k\hbar
\omega_{Dk} b_k^{\dagger} b_k +(U_c+\frac{\chi^2}{\hbar \omega_o})
\hat{n}^2
\nonumber \\
+ \sum_k T_{Sk} \left (a_k c^\dagger e^{\lambda (a^\dagger -a)
}+ c a_k^\dagger e^{-\lambda (a^\dagger -a)}\right ) \nonumber \\
+ \sum_kT_{Dk} \left (b_k c^\dagger e^{\lambda
(a^\dagger -a)} + c b_k^\dagger e^{-\lambda
(a^\dagger -a)} \right )\; .
\end{eqnarray}
We will now work exclusively in this displacement picture.

To derive a master equation for the dot, we first transform to an
interaction picture in the usual way to give the Hamiltonian
\begin{eqnarray}
H_{I} &=&\sum T_{Sk}(a_{k}c^{\dagger }e^{i(\omega _{I}-\eta
-\omega _{Sk})t}e^{-\lambda(a^{\dagger}e^{i\omega
_{o}t}-ae^{-i\omega _{o}t})}
\nonumber \\
&&\hspace{0.5in}+ca_{k}^{\dagger }e^{-i(\omega _{I}-\eta
-\omega _{Sk})t}e^{\lambda(a^{\dagger
}e^{i\omega
_{o}t}-ae^{-i\omega _{o}t})}) \nonumber \\
&&+\sum T_{Dk}(b_{k}c^{\dagger }e^{i(\omega _{I}-\eta -\omega _{Dk})t}e^{-%
\lambda(a^{\dagger }e^{i\omega
_{o}t}-ae^{-i\omega
_{o}t})} \nonumber \\
&&\hspace{0.5in}+cb_{k}^{\dagger }e^{-i(\omega _{I}-\eta
-\omega _{Dk})t}e^{\lambda(a^{\dagger
}e^{i\omega _{o}t}-ae^{-i\omega _{o}t})}),
\end{eqnarray}
where $\eta = {\chi^2}/({\hbar \omega_o})=\chi\lambda$.

At this point we might wish to trace out the phonon bath, however
we will postpone this  for a closer look at the tunneling
Hamiltonian at the individual phonon level. We use the exponential
approximation $e^x= 1 + x + x^2/(2!)+\cdots$ , when $x$ is small
for the term $e^{\lambda (a^\dagger e^{i \omega_o t} - a e^{-i
\omega_o t})}$. We expect an expansion to second order in
$\lambda$ to give an adequate description of transport,  in that
at least one step in the current vs.~bias voltage curve is seen due to
phonon mediated
tunneling.   In the experiment of Park et al., $\lambda$ was less
than unity, but not very small. Strong coupling between the
electronic and vibrational degrees of freedom (large $\lambda$)
will give multi-phonon tunneling events, and corresponding
multiple steps in the current vs.~bias voltage curves. The
Hamiltonian can then be written as
\begin{eqnarray}
H_{I} &=& \sum T_{Sk}(a_{k}c^{\dagger }e^{i(\omega _{I}-\eta
-\omega
_{Sk})t}+ca_{k}^{\dagger }e^{-i(\omega _{I}-\eta -\omega _{Sk})t})
\label{e:Hinter}\nonumber \\
&&+ \lambda \sum T_{Sk}(a_{k}c^{\dagger }a e^{i(\omega
_{I}-\eta -\omega _{Sk}-\omega _{o})t}+ca_{k}^{\dagger
}a^{\dagger }e^{-i(\omega _{I}-\eta
-\omega _{Sk}-\omega _{o})t} \nonumber \\
&&\hspace{0.5in}-a_{k}c^{\dagger }a^{\dagger }e^{i(\omega
_{I}-\eta -\omega _{Sk}+\omega _{o})t}-ca_{k}^{\dagger
}ae^{-i(\omega _{I}-\eta -\omega _{Sk}+\omega
_{o})t}) \nonumber \\
&&+\frac{\lambda^2}{2} \sum T_{Sk}(a_{k}c^{\dagger }a a e^{i(\omega
_{I}-\eta -\omega _{Sk}-2\omega _{o})t}+ca_{k}^{\dagger
}a^{\dagger }a^{\dagger }e^{-i(\omega _{I}-\eta
-\omega _{Sk}-2\omega _{o})t} \nonumber \\
&&\hspace{0.5in}-2a_{k}c^{\dagger }a^{\dagger }a e^{i(\omega
_{I}-\eta -\omega _{Sk})t}-2 ca_{k}^{\dagger }a^{\dagger} a
e^{-i(\omega _{I}-\eta -\omega _{Sk})t} \nonumber \\
&&\hspace{0.5in}+a_{k}c^{\dagger }a^{\dagger }a^{\dagger
}e^{i(\omega _{I}-\eta -\omega _{Sk}+2\omega
_{o})t}-ca_{k}^{\dagger }aae^{-i(\omega _{I}-\eta -\omega
_{Sk}+2\omega_{o})t}) \nonumber \\
&&+\sum T_{Dk}(b_{k}c^{\dagger }e^{i(\omega _{I}-\eta -\omega
_{Dk})t}+cb_{k}^{\dagger }e^{-i(\omega _{I}-\eta -\omega _{Dk})t})
\nonumber \\
&&+\lambda \sum T_{Dk}(-b_{k}c^{\dagger }a^{\dagger }e^{i(\omega
_{I}-\eta -\omega _{Dk}+\omega _{o})t}-cb_{k}^{\dagger
}ae^{-i(\omega _{I}-\eta -\omega _{Dk}+\omega _{o})t} \nonumber \\
&&\hspace{0.5in}+b_{k}c^{\dagger }ae^{i(\omega _{I}-\eta
-\omega _{Dk}-\omega _{o})t}+cb_{k}^{\dagger }a^{\dagger
}e^{-i(\omega _{I}-\eta -\omega _{Dk}-\omega _{o})t})
\nonumber\\
&&+\frac{\lambda^2}{2}\sum T_{Dk}(b_{k}c^{\dagger }a a e^{i(\omega
_{I}-\eta -\omega_{Dk}-2\omega _{o})t}+c b_{k}^{\dagger
}a^{\dagger }a^{\dagger }e^{-i(\omega _{I}-\eta
-\omega _{Dk}-2\omega _{o})t} \nonumber \\
&&\hspace{0.5in}-2b_{k}c^{\dagger }a^{\dagger }a e^{i(\omega
_{I}-\eta -\omega _{Dk})t}-2 cb_{k}^{\dagger }a^{\dagger} a
e^{-i(\omega _{I}-\eta -\omega _{Dk})t} \nonumber \\
&&\hspace{0.5in}+b_{k}c^{\dagger }a^{\dagger }a^{\dagger
}e^{i(\omega _{I}-\eta -\omega _{Dk}+2\omega
_{o})t}-cb_{k}^{\dagger }aae^{-i(\omega _{I}-\eta -\omega
_{Dk}+2\omega_{o})t}).
\end{eqnarray}
The terms of zero order in $\lambda$ describe bare tunneling
through the system and do not cause excitations of the vibrational
degree of freedom. The terms linear in $\lambda$ correspond to the
exchange of one vibrational quantum, or phonon. The terms
quadratic in $\lambda$ correspond to tunneling with the exchange
of two vibrational quanta. Higher order terms could obviously be
included at considerable computational expense. We will proceed to
derive the master equation  up to quadratic order in $\lambda$.

\section{Master equation}
Our objective here is to find an evolution equation of the joint  density
operator for the  electronic and vibrational degrees of freedom.
 We will use standard methods based on the Born
and Markov approximation\cite{Gardiner-Zoller}.
In order to indicate where these approximations occur, we will sketch some of
the key elements of the derivation in what follows.
The Born approximation assumes that the coupling between the leads and
the local system is weak and thus second order perturbation theory
will suffice to describe this interaction;
\begin{eqnarray}
\dot{\rho} &=& \frac{-1}{\hbar^2} \int_0^t dt' \rm{Tr}
[H_I(t),[H_I(t'),R]],
\label{e:mastereqngen}
\end{eqnarray}
where $R$ is the joint density matrix for the vibrational and
electronic degrees of freedom of the local system and the reservoirs.

At this point we would like to trace out the electronic degrees of
freedom for the source and drain.  We will assume that the states
of the source and drain remain in local thermodynamic equilibrium
at temperature $T$. This is part of the Markov approximation. Its
validity requires that any correlation that develops between the
electrons in the leads and the local system, as a result of the
tunneling interaction, is rapidly damped to zero on time scales
relevant for the system dynamics.  We need the following moments:
\begin{eqnarray}
{\rm Tr}[a_{k}^\dagger a_{k} \rho] = f_{Sk}, \hspace{2cm}
&& {\rm Tr}[b_{k}^\dagger b_{k} \rho] = f_{Dk}, \nonumber \\
{\rm Tr}[a_{k} a_{k}^\dagger \rho] = 1 - f_{Sk}, \hspace{2cm} &&
{\rm Tr}[b_{k} b_{k}^\dagger \rho] = 1 - f_{Dk}. \nonumber
\end{eqnarray}
where $f_{Sk}=f(E_{Sk})$ is the Fermi function describing the
average occupation number in the source and similarly
$f_{Dk}=f(E_{Dk})$, for the drain.
The Fermi function has an implicit dependence on the temperature, $T$,
of the electronic system.

The next step is to convert the sum over modes to a
frequency-space integral:
\begin{eqnarray} \sum_k f_{Sk} |T_{Sk}|^2 \rightarrow \int_0^\infty
d\omega \hspace{1mm} g(\omega) f_D(\omega) | T_S(\omega) |^2,
\end{eqnarray}
where $|T_{Sk}|^2=T_{Sk}^*T_{Sk}$ and $g(\omega)$ is the
density of states.
 Evaluating the time integral, we use,
\begin{eqnarray}
\int_0^\infty d\tau e^{\pm i \epsilon \tau} = \pi \delta(\epsilon)
\pm i PV(1/\epsilon),
\end{eqnarray}
where $\tau = t - t'$ and the imaginary term is ignored.

Using these methods, we can combine the terms
for the source and drain as the left and right tunneling rates,
$\gamma_L$ and $\gamma_R$
respectively
\begin{eqnarray}
\int_0^\infty d\omega \hspace{1mm} g(\omega) |T_S(\omega)|^2
\delta(\omega_0) = \gamma_L(\omega_0).
\end{eqnarray}
In the same way, we can define
\begin{eqnarray} \gamma_{L1} &=& \gamma_L
(\hbar \omega_I-\eta-\mu_L), \nonumber \\
f_{1L} &=& f(\hbar \omega_I-\eta - \mu_L), \nonumber \\
\gamma_{L2} &=& \gamma_L (\hbar \omega_I -\eta
- \hbar \omega_o - \mu_L),
\nonumber \\
f_{2L} &=& f(\hbar \omega_I-\eta - \hbar \omega_o -\mu_L), \nonumber \\
\gamma_{L3} &=& \gamma_L (\hbar \omega_I -\eta
+ \hbar \omega_o -\mu_L), \nonumber \\
f_{3L} &=& f(\hbar \omega_I-\eta + \hbar \omega_o -\mu_L).
\nonumber
\end{eqnarray}
and similarly for $\gamma_{R1},\gamma_{R2} \gamma_{R3}, f_{1R},
f_{2R}, f_{3R}$ replacing $\mu_L$ with $\mu_R$ and $f$ being the
Fermi functions which have a dependence on the bias voltage (through
the chemical potential) and also on the phonon energy
$\hbar\omega_o$. As the bias voltage is increased from zero, the
first Fermi function to be significantly different from zero is
$f_{2L}$  followed by $f_{1L}$ and then $f_{3L}$.  This stepwise
behavior will be important in understanding the dependence of the
stationary current as a function of bias voltage.

The master equation in the canonical transformed picture to the
second order in $\lambda$ may be written as
\begin{eqnarray}
\frac{d\bar{\rho}}{dt}&=&   \gamma_{L1} \biggl( (1-\lambda^2)
(f_{1L}\mathcal{D}
[c^\dagger]\bar{\rho}+(1-f_{1L})\mathcal{D} [c]\bar{\rho} )  \label{e:mastereqnsimple}\nonumber \\
&&\hspace{0.3in}+  \lambda^2\ (f_{1L} (- a^{\dagger}a c^\dagger
\bar{\rho} c
+ a^\dagger a c c^\dagger \bar{\rho} - c^\dagger \bar{\rho} c a^\dagger a + \bar{\rho} c c^\dagger a^\dagger a) \nonumber\\
&&\hspace{0.5in}+ (1-f_{1L}) (- a^{\dagger}a c \bar{\rho}
c^\dagger + a^\dagger a c^\dagger c \bar{\rho}
- c \bar{\rho} c^\dagger a^\dagger a + \bar{\rho} a^\dagger a c^\dagger c ) ) \biggr) \nonumber\\
&+&\gamma _{L2} \lambda ^2 \biggl(f_{2L} \mathcal{D}[a c^\dagger]
\bar{\rho} +(1-f_{2L})\mathcal{D}[a^\dagger c]\bar{\rho} \biggr) \nonumber \\
&+&\gamma _{L3} \lambda ^2 \biggl(f_{3L} \mathcal{D}[a^\dagger
c^\dagger]\bar{\rho} +(1-f_{3L})\mathcal{D}[ac]\bar{\rho} \biggr) \nonumber \\
&+& \gamma _{R1} \biggl((1-\lambda^2) (f_{1R} \mathcal{D} [c^\dagger]\bar{\rho} +(1-f_{1R}) \mathcal{D}[c]\bar{\rho}) \nonumber \\
&&\hspace{0.3in}+ \lambda^2\ (f_{1R} (- a^{\dagger}a c^\dagger
\bar{\rho} c
+ a^\dagger a c c^\dagger \bar{\rho} - c^\dagger \bar{\rho} c a^\dagger a + \bar{\rho} c c^\dagger a^\dagger a) \nonumber\\
&&\hspace{0.5in}+(1-f_{1R}) (- a^{\dagger}a c \bar{\rho} c^\dagger
+ a^\dagger a c^\dagger c \bar{\rho}
- c \bar{\rho} c^\dagger a^\dagger a + \bar{\rho} a^\dagger a c^\dagger c ) ) \biggr) \nonumber\\
&+&\gamma _{R2} \lambda^2 \biggl(f_{2R} \mathcal{D}[a c^\dagger]
\bar{\rho} +(1-f_{2R}) \mathcal{D}[a^\dagger c] \bar{\rho} \biggr) \nonumber \\
&+&\gamma _{R3} \lambda^2\biggl(f_{3R} \mathcal{D} [a^\dagger
c^\dagger] \bar{\rho} +(1-f_{3R}) \mathcal{D}[a c]\bar{\rho} \biggr) \nonumber \\
&+& \kappa (\bar{n}_p +1) \mathcal{D}[a] \bar{\rho} + \kappa
\bar{n}_p \mathcal{D}[a^\dagger] \bar{\rho} + \kappa \lambda^2 (2
\bar{n}_p +1) \mathcal{D}[c^\dagger c] \bar{\rho}\; ,
\end{eqnarray}
where the notation ${\cal D}$ is defined for arbitrary operators $X$
and $Y$ as
\begin{eqnarray}
\mathcal{D}[X]Y &=& \mathcal{J}[X]Y - \mathcal{A}[Y] \nonumber \\
&=& X Y X^\dagger -\frac{1}{2} (X^\dagger X Y + Y X^\dagger X),
\end{eqnarray}
and
\begin{equation}
\bar{n}_p(\omega_o) = \frac{1}{e^{\hbar \omega_o/k_BT}-1} \;.
\end{equation}
We have included in this model an explicit damping process of the
oscillators motion at rate $\kappa$ into a thermal oscillator bath
with mean excitation $\bar{n}_p$ and $T$ is the effective temperature
of reservoir responsible for this damping process.  A possible
physical origin for this kind of damping could be as follows. Thermal
fluctuations in the metal contacts of the source and drain cause
fluctuations in position of the center of the trapping potential
confining the molecule, that is to say small, fluctuating linear
forces act on the molecule.  For a harmonic trap, this appears to the
oscillator as a thermal bath. However we expect such a mechanism to be
very weak. This fact, together with the very large frequency of the
oscillator, justifies our use of the quantum optical master equation
(as opposed to the Brownian motion master equation)  to describe this
source of dissipation\cite{Gardiner-Zoller}. The thermo-mechanical and
electronic temperatures are not necessarily the same, although we will
generally assume this to be the case.

Setting $\lambda=0$ we recover the standard master equation for a
single quantum dot coupled to leads\cite{Sun}. The superoperator
${\cal D}[c^\dagger]$ adds one electron to the dot.   Terms containing
this super operator describe a conditional Poisson event in which an
electron tunnels onto the dot. The  electron can enter from the
source, with probability per unit time of $\gamma_{L1}f_{1L}\langle
cc^\dagger\rangle$, or it can enter from the drain, with probability
per unit time $\gamma_{R1}f_{1R}\langle cc^\dagger\rangle$.  Likewise
the term ${\cal D}[c]$ describes an electron leaving the dot, again
via tunneling into the source (terms proportional to $\gamma_{L1}$)
or the drain  (terms proportional to $\gamma_{R1}$). When $\lambda\neq
0$ there are additional terms describing phonon mediated tunneling
events onto and off the dot. Any term proportional to $\gamma_{Li},\
i=1,2,3.$ describes an electron tunneling out of, or into, the
source, while any term proportional to $\gamma_{Ri},\ i=1,2,3$
describes an electron tunneling out of, or into, the drain.

The average currents through the left junction (source lead--dot) and the
right junction (dot--drain lead) are related  to each other,
and the average occupation of the dot, by
\begin{equation}
I_L(t)-I_R(t)=e\frac{d\langle c^\dagger c\rangle}{dt}.
\label{current}
\end{equation}
In the steady state, the occupation of the dot is constant and the
average currents through the two junctions are
equal. Of course, the actual
fluctuating time dependent currents are almost never equal. The
external current arises as the external circuit adjusts the chemical
potential of the local Fermi reservoir when electrons tunnel onto or
off the dot. It is thus clear that the current through the left junction must
depend only on the tunneling rates $\gamma_{Li},\ i=1,2,3.$ in the
left barrier. This makes it easy to identify the average current through
the left (or right) junction by inspection of the equation of motion for
$\langle c^\dagger c\rangle$: all terms in the right hand side of
Eq.~(\ref{current}) proportional to $\gamma_{Li}$ correspond to the
average current through the left junction, $I_L(t)$, while all terms on the
right hand side proportional to $\gamma_{Ri}$ correspond to the
negative of the current through the right junction, $-I_R(t)$.

\section{Local system dynamics}
We can now compute the current through the quantum dot.  The
current reflects how the reservoirs of the source and drain
respond to the dynamics of the vibrational and electronic degrees
of freedom. Of course in an experiment the external current is
typically all we have access to. However  the master equation
enables us to calculate the coupled dynamics of the vibrational
and electronic degrees of freedom. Understanding this dynamics is
crucial to explaining the observed features in the external
current. As electrons tunnel on and off the dot, the oscillator
experiences a force due to the electrostatic potential. While the
force is conservative, the tunnel events are stochastic (in fact a
Poisson process) and thus the excitation of the oscillator is
stochastic. Furthermore the vibrational and electronic degrees of
freedom become correlated through the dynamics. In this section we
wish to investigate these features in some detail.

From the master equation, the rate of change of this average
electron number in the dot may be obtained:
\begin{eqnarray}
 \frac{d\langle c^\dagger c \rangle _{\rm CT}}{dt} &=& {\rm tr}[c^\dagger c\frac{d\bar{\rho}}{dt}] \\
&=& [\gamma_{L1} (1- \lambda^2) (f_{1L}-\langle c^\dagger c \rangle)+\gamma_{R1} (1- \lambda^2) ( f_{1R} -\langle c^\dagger c\rangle)\nonumber\\
 & & -2\gamma_{L1}\lambda^2(f_{1L}\langle a^\dagger a\rangle-\langle a^\dagger a c^\dagger c\rangle)\nonumber\\
 & & +\gamma_{L2}\lambda^2(f_{2L}\langle a^\dagger a\rangle-\langle a^\dagger a c^\dagger c\rangle-(1-f_{2L})\langle c^\dagger c\rangle)\nonumber\\
 & & +\gamma_{L3}\lambda^2(f_{3L}\langle 1+a^\dagger a\rangle-f_{3L}\langle c^\dagger c\rangle-\langle a^\dagger a c^\dagger c \rangle)\nonumber\\
  &  & -2\gamma_{R1}\lambda^2(f_{1R}\langle a^\dagger a\rangle-\langle a^\dagger a c^\dagger c\rangle)\nonumber\\
 & & +\gamma_{R2}\lambda^2(f_{2R}\langle a^\dagger a\rangle-\langle a^\dagger a c^\dagger c\rangle-(1-f_{2R})\langle c^\dagger c\rangle)\nonumber\\
 & & +\gamma_{R3}\lambda^2(f_{3R}\langle 1+a^\dagger
 a\rangle-f_{3R}\langle c^\dagger c\rangle-\langle a^\dagger a
 c^\dagger c \rangle)]_{\rm CT} \ .
 \label{eq:cdc_CT}
 \end{eqnarray}

While for the vibrational degrees of freedom, we see that
\begin{eqnarray}
 \frac{d\langle a^\dagger a \rangle_{\rm CT}}{dt} &=& {\rm tr}[a^\dagger a
\frac{d\bar{\rho}}{dt}] \\
&=& \lambda^2 [\gamma_{L2}(-f_{2L}\langle a^\dagger a\rangle+\langle a^\dagger a c^\dagger c\rangle+(1-f_{2L})\langle c^\dagger c\rangle) \nonumber \\
& & +\gamma_{L3}(f_{3L}\langle 1+a^\dagger a\rangle-f_{3L}\langle c^\dagger c\rangle-\langle a^\dagger a c^\dagger c\rangle) \nonumber \\
& & +\gamma_{R2}(-f_{2R}\langle a^\dagger a\rangle+\langle a^\dagger a c^\dagger c\rangle+(1-f_{2R})\langle c^\dagger c\rangle) \nonumber \\
& & +\gamma_{R3}(f_{3R}\langle 1+a^\dagger a\rangle-f_{3R}\langle
c^\dagger c\rangle-\langle a^\dagger a c^\dagger
c\rangle)-\kappa\langle a^\dagger a\rangle]_{\rm CT}+\kappa\bar{n}_p \ .
\label{eq:ada_CT}
\end{eqnarray}
Here the subscript CT indicates that the quantity to which it is
attached is evaluated in the canonical transformed (CT) basis. The
average occupational number of electron in the dot in the original
basis is the same as in the CT basis:
\begin{equation}
\langle c^\dagger c \rangle = {\rm tr} [c^\dagger c \rho]= {\rm tr} [c^\dagger c \bar{\rho}]=\langle c^\dagger c \rangle _{\rm CT}. \nonumber
\end{equation}
While for the vibrational degrees of freedom, we have
\begin{eqnarray}
\langle a^\dagger a \rangle &=& {\rm tr} [a^\dagger a \rho]
\nonumber \\
&=& {\rm tr} [a^\dagger a e^{-s} e^{-i \omega_o a^\dagger a t} \bar{\rho} e^{i \omega_o a^\dagger a t} e^{s}] \nonumber \\
&=& {\rm tr} [e^{i \omega_o a^\dagger a t}(a^\dagger + \lambda \hat{n}) (a +\lambda \hat{n}) e^{-i \omega_o a^\dagger a t} \bar{\rho}] \nonumber \\
&=& \langle a^\dagger a \rangle_{\rm CT} + \lambda \langle
(a^\dagger e^{-i \omega_o a^\dagger a t}+ a e^{-i \omega_o
a^\dagger a t}) \hat{n} \rangle_{\rm CT}+ \lambda^2 \langle
\hat{n}^2 \rangle.
\label{CTadaTr}
\end{eqnarray}
If the initial displacement  $\langle x\rangle $ is zero, the second
time dependent term in the previous expression remains zero. We will
assume this is the case in what follows.

In general we do not get a closed set of equations for the mean
phonon and electron number due to the presence in these equations
of the  higher order moment $\langle a^\dagger a c^\dagger
c\rangle$.  This reflects the fact that the electron and
vibrational degrees of freedom are correlated (and possibly
entangled) through the dynamics. One might  proceed by introducing
a semiclassical factorization approximation by  replacing $\langle
a^\dagger a c^\dagger c \rangle$ by the factorized average values,
i.e., $\langle a^\dagger a c^\dagger c \rangle\approx \langle
a^\dagger a \rangle \langle c^\dagger c \rangle$, then the
evolution equations (\ref{eq:cdc_CT}) and (\ref{eq:ada_CT}) forms
a closed set of equations.  However there is a special case for
which this is not necessary. If we let $\gamma_{L1} = \gamma_{L2}
= \gamma_{L3} = \gamma_L $ and $\gamma_{R1} = \gamma_{R2} =
\gamma_{R3} = \gamma_R $ which is the assumption of
energy-independent tunnel couplings, the equations do form a
closed set:
\begin{eqnarray}
 \frac{d\langle c^\dagger c \rangle}{dt} &=& A_1 \langle c^\dagger
 c \rangle + B_1 \langle a^\dagger a \rangle _{\rm CT} + C_1 \; ,
\label{e:cdagcevol}\\
 \nonumber \\
 A_1 &=& - \bigl[\gamma_L(1 - f_{2L} \lambda^2 + f_{3L} \lambda^2)+ \gamma_R(1 - f_{2R} \lambda^2+ f_{3R}
 \lambda^2)\bigr], \nonumber \\
 B_1 &=& \lambda^2 (-2 f_{1L} \gamma_L + f_{2L} \gamma_L +f_{3L} \gamma_L -2 f_{1R} \gamma_R + f_{2R}
 \gamma_R +f_{3R} \gamma_R), \nonumber \\
C_1 &=& (1-\lambda^2) \gamma_L f_{1L} + \gamma_L f_{3L} \lambda^2
+ (1-\lambda^2) \gamma_R f_{1R} + \gamma_R f_{3R} \lambda^2 \;
,\nonumber
 \end{eqnarray}
and
\begin{eqnarray}
 \frac{d\langle a^\dagger a \rangle_{\rm CT}}{dt} &=& A_2 \langle c^\dagger
 c \rangle + B_2 \langle a^\dagger a \rangle _{\rm CT}+ C_2 \; ,
\label{e:adagaevol}\\\
 \nonumber\\
  A_2 &=&  \lambda^2 (\gamma_L ( 1 - f_{2L} - f_{3L}) + \gamma_R (1 -
  f_{2R} - f_{3R})),  \nonumber
 \\
 B_2 &=&  \lambda^2 (-\gamma_L f_{2L} + \gamma_L f_{3L} - \gamma_R f_{2R} +
 \gamma_R f_{3R}) - \kappa , \nonumber \\
 C_2 &=&  \lambda^2 (\gamma_L f_{3L} + \gamma_R f_{3R}) + \kappa
 \bar{n}_p\; . \nonumber
\end{eqnarray}
Consideration of Eq.~(\ref{e:adagaevol}) indicates that it is possible for the
oscillator to achieve a steady state even when there is no explicit
thermo-mechanical damping ($\kappa=0$).  This requires bias conditions
such that $f_{3L}=f_{3R}=0$. It is remarkable that the process of
electrical transport between source and drain alone can induce damping
of the mechanical motion. This result has been indicated by other
authors\cite{Mozyrsky-Martin,Smirnov,Mozyrsky-Martin-Hastings,Armour}.

\begin{figure}
\unitlength1cm
\begin{minipage}[t]{8cm}
\includegraphics[width=7cm]{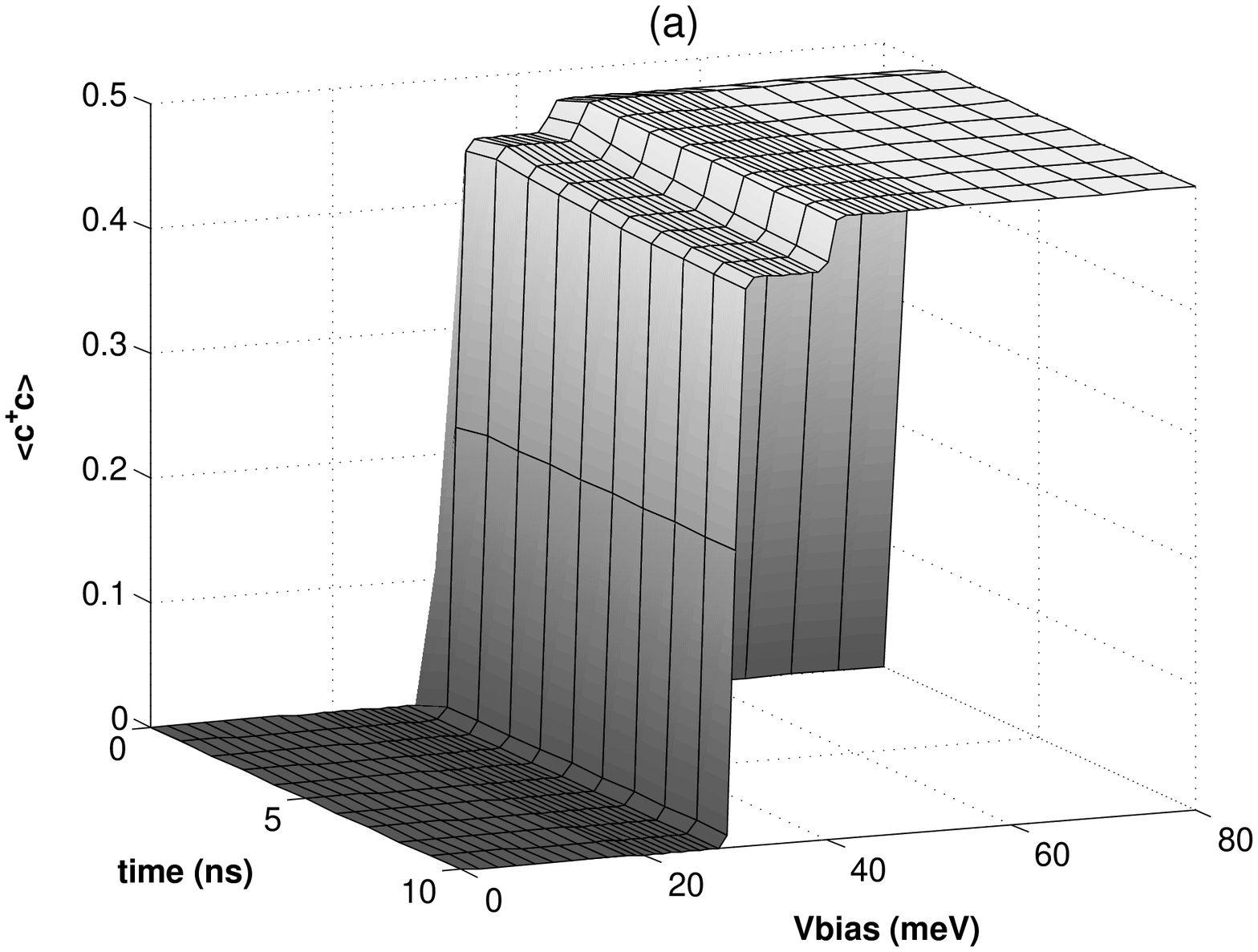}
\end{minipage}
\begin{minipage}[t]{8cm}
\includegraphics[width=7cm]{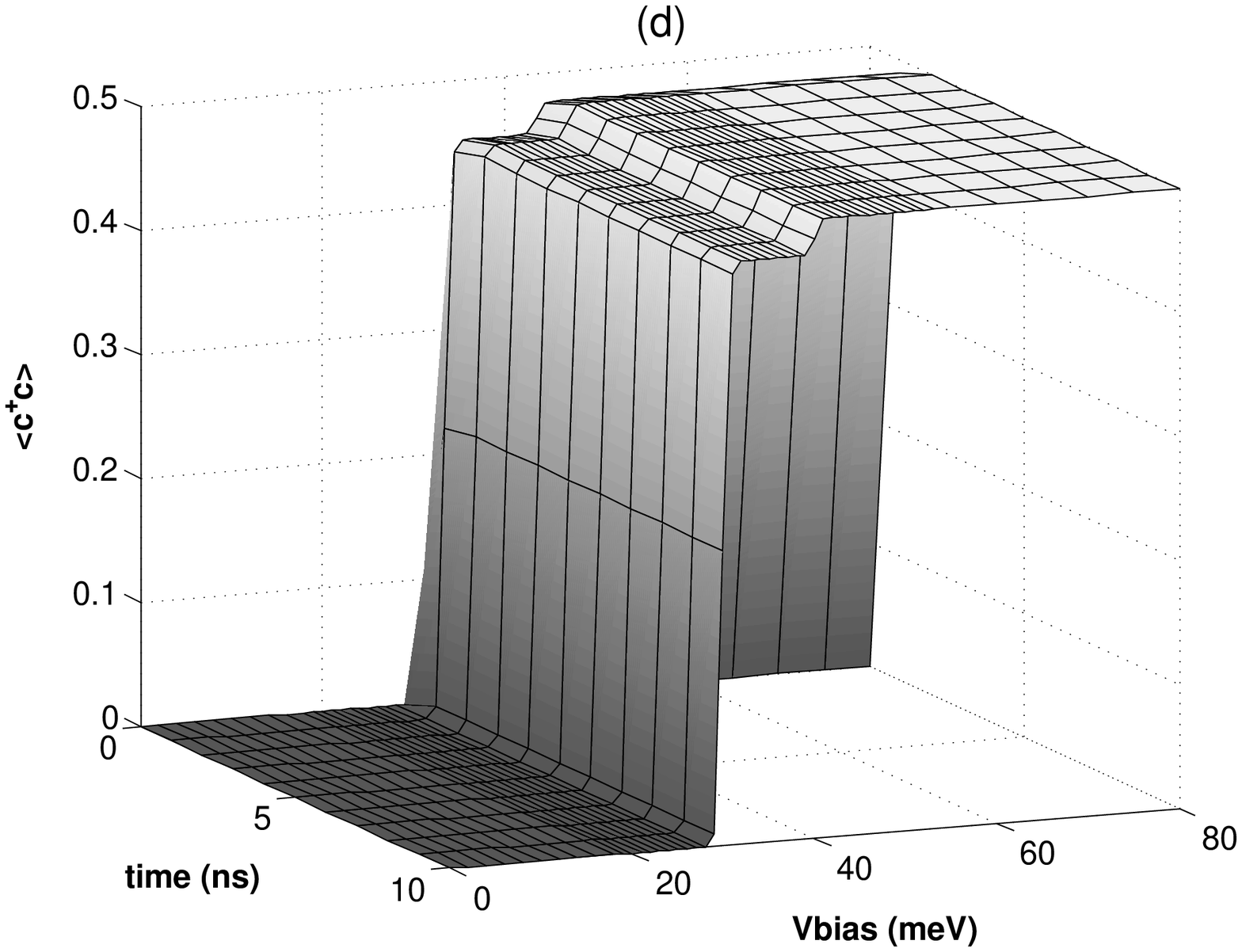}
\end{minipage}
\begin{minipage}[t]{8cm}
\includegraphics[width=7cm]{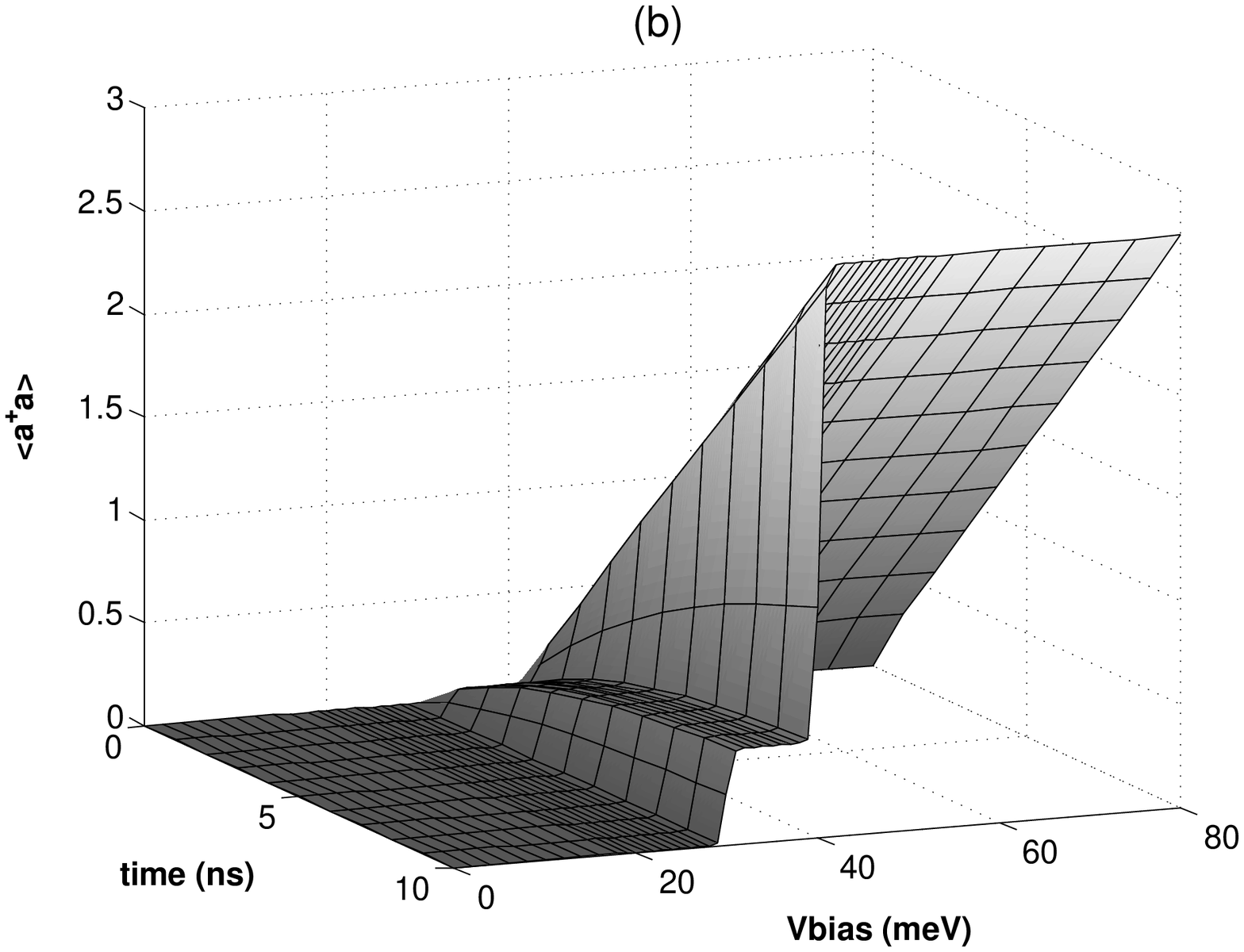}
\end{minipage}
\unitlength1cm
\begin{minipage}[t]{8cm}
\includegraphics[width=7cm]{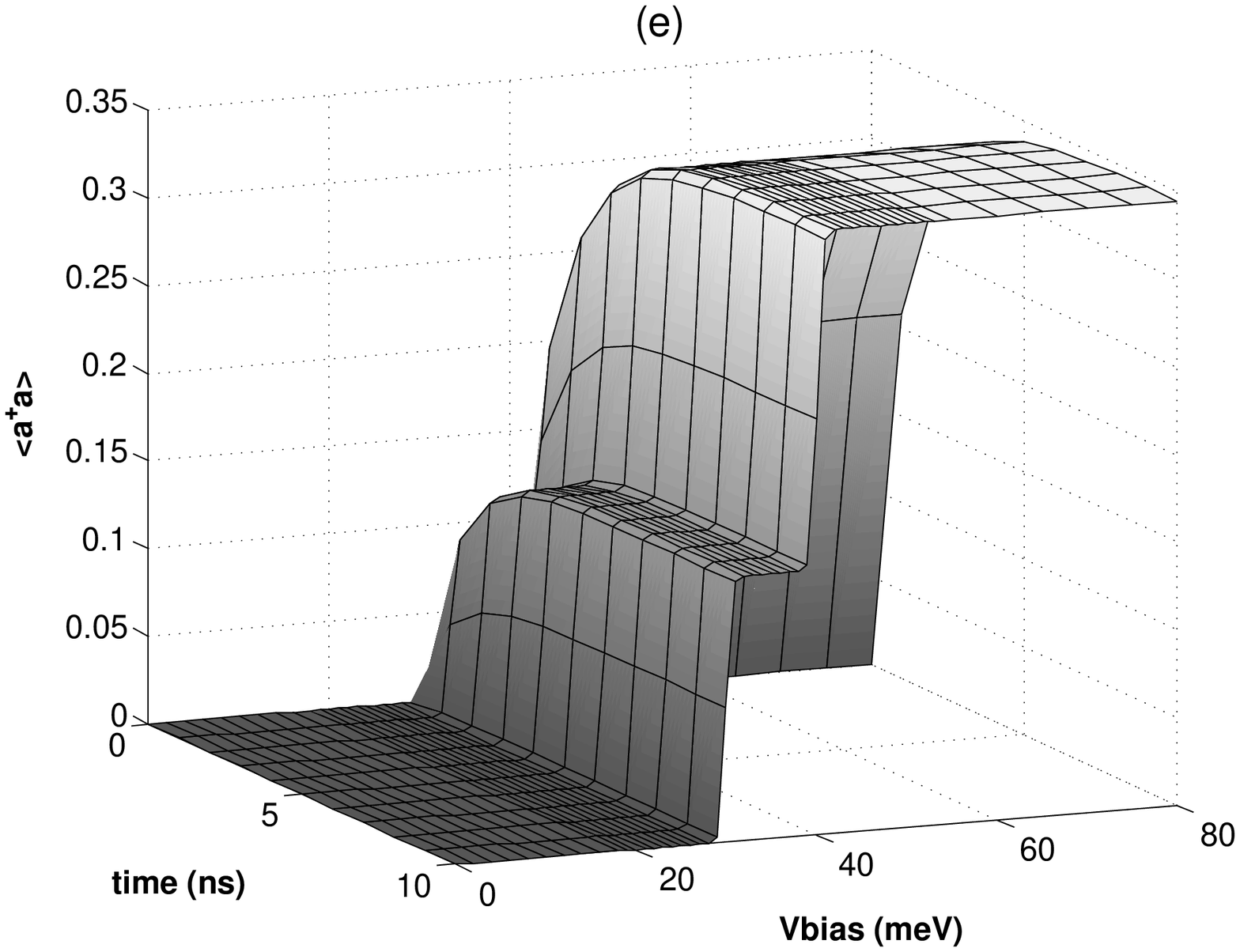}
\end{minipage}
\begin{minipage}[t]{8cm}
\includegraphics[width=7cm]{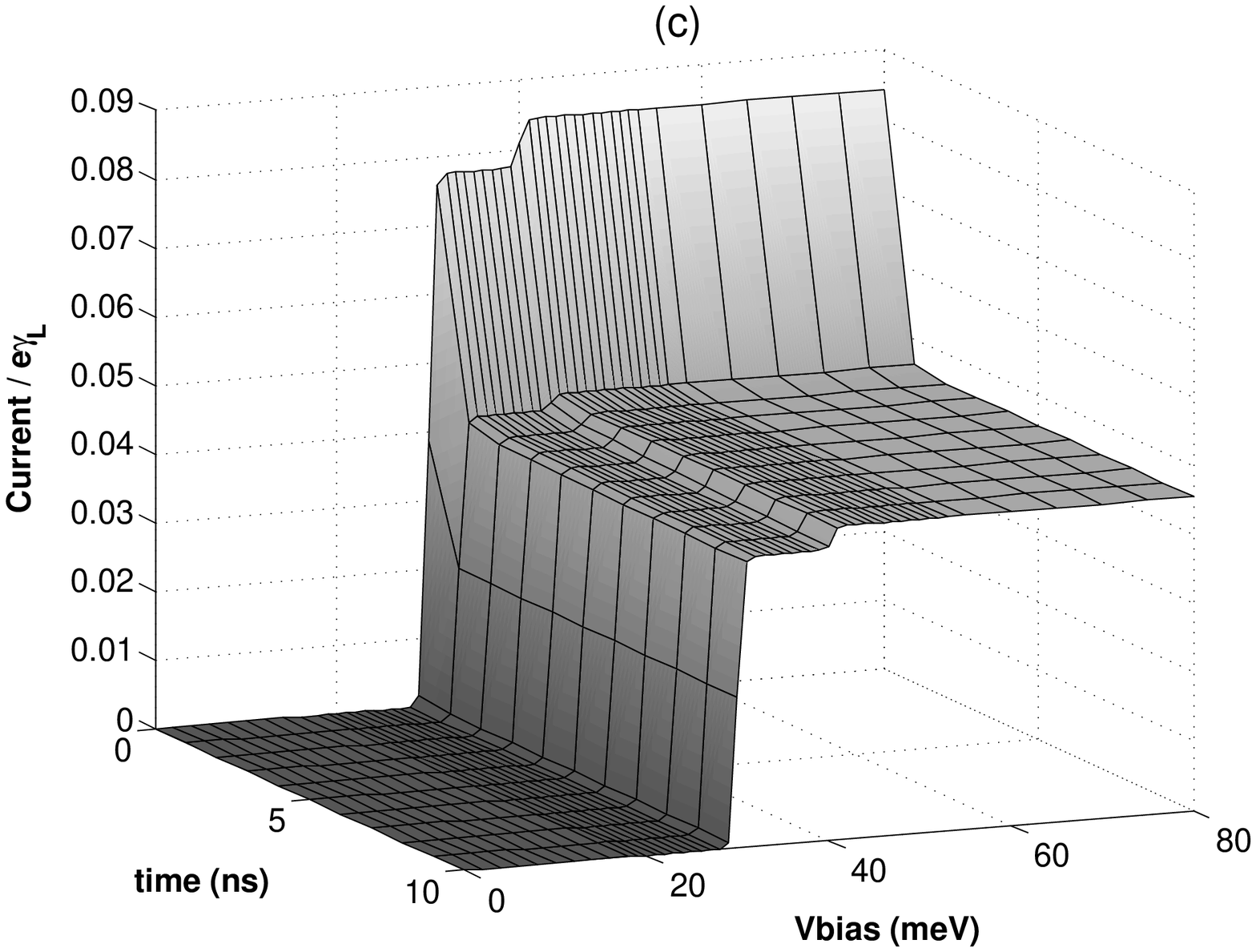}
\end{minipage}
\begin{minipage}[t]{8cm}
\includegraphics[width=7cm]{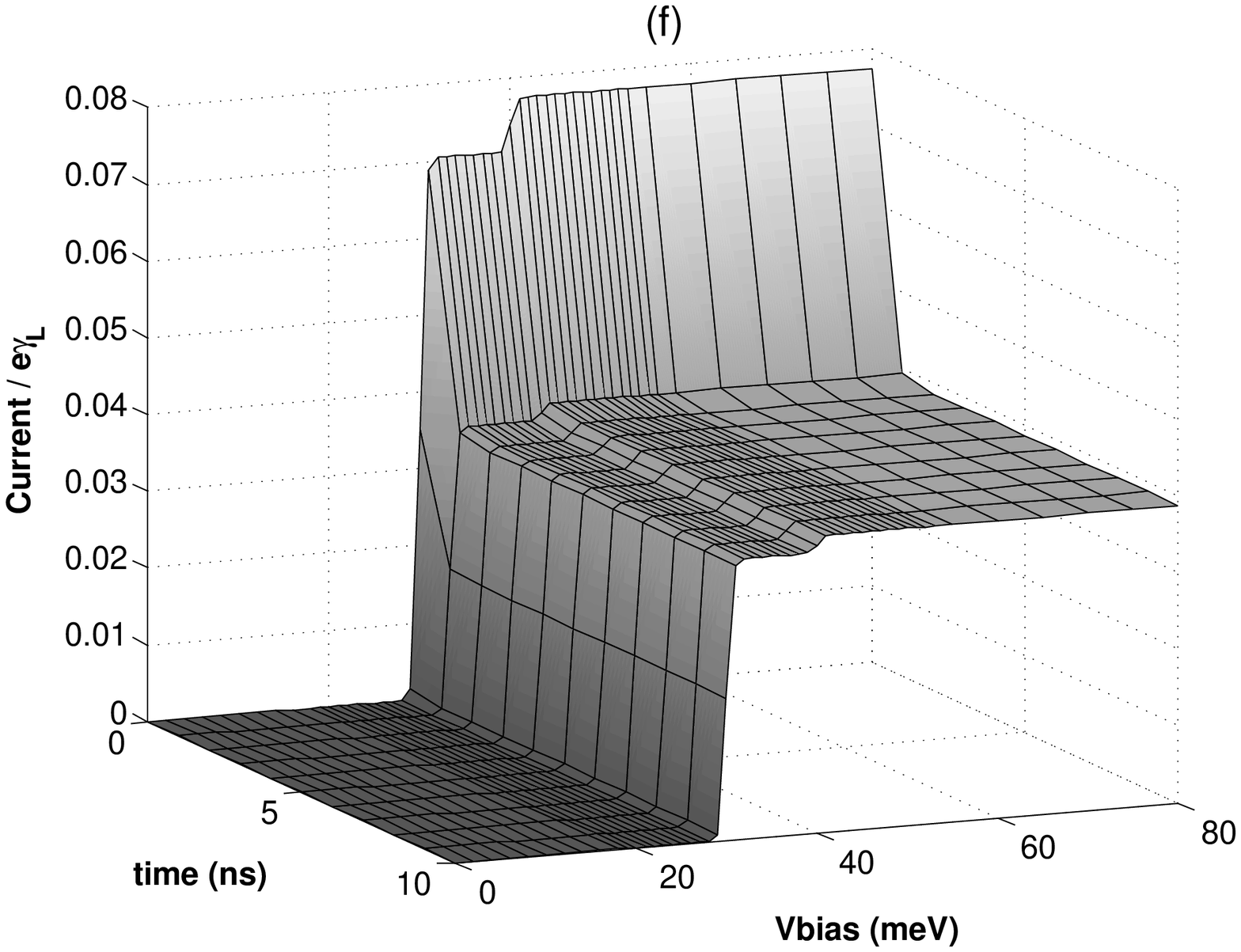}
\end{minipage}
\unitlength1cm
\caption{Average number of electron, phonon and current through
the dot against bias voltage with $\hbar \omega_o =$ 5 meV, $\hbar
\omega_I -\eta$ = 15 meV, $k_B T =$ 0.13 meV, and $\hbar \gamma_L =
\hbar \gamma_R =$ 2 $\mu$eV for $\lambda = 0.3$. Figures
(a),(b),(c) without damping and (d),(e),(f) with damping $\kappa =
0.3 \gamma_L$. We assume $\mu_{L}=-\mu_{R}=eV_{\rm bias}/2$. \label{fig2}}
\end{figure}

These equations were solved numerically  and the results, for
various values of $\lambda$ and bias voltage, are shown in
Fig.~\ref{fig2}.  A feature of our approach is that we can directly
calculate the dynamics of the local degrees of freedom, for
example the mean electron occupation of the dot as well as the
mean vibrational occupation number in the oscillator.

From these equations we can reproduce behavior for the stationary
current similar to that observed in the experiment. We concentrate
here on the stationary current through the left junction (connected to the
source). Similar results apply for the right junction. We assume that
the electronic temperature is $1.5$K, which is the temperature
used in the experiment by Park et al. \cite{Park}.

Following the discussion below Eq.~(\ref{current}),
%at the start of this section,
we see from Eq.~(\ref{e:cdagcevol}) that the average steady state
current through the left junction is given by
\begin{eqnarray}
I_{\rm st} &=& e\gamma_L [(-1+ f_{2L}\lambda^2 - f_{3L}\lambda^2)
\langle c^\dagger c \rangle_{\rm st} +(-2 f_{1L}\lambda^2 + f_{2L}\lambda^2
+ f_{3L}\lambda^2) \langle a^\dagger a \rangle_{\rm CT, st}\nonumber \\
& &  +(1-\lambda^2)f_{1L}+f_{3L}\lambda^2].
\label{Ist}
\end{eqnarray}
which is of course equal to the average steady state current through the
right junction. The steady state current, $I_{\rm st}$ can then be found
by finding the steady state
solution for each of the phonon number and electron number,
\begin{eqnarray}
\langle c^\dagger c \rangle_{\rm st} &=& \frac{B_1 C_2 - B_2
C_1}{A_1 B_2 - A_2 B_1}\; ,\\
\langle a^\dagger a \rangle_{\rm CT, st} &=& \frac{-A_2}{B_2}
\biggl( \frac{B_1 C_2 - B_2 C_1}{A_1 B_2 - A_2 B_1}\biggr) -
\frac{C_2}{B_2}\; .
\end{eqnarray}

In Fig.~\ref{fig2}, we assume that the bias voltage is applied
symmetrically, i.e, $\mu_{L}=-\mu_{R}=eV_{\rm bias}/2$. In this
case, all the Fermi factors $f_{1R}$, $f_{2R}$, and $f_{3R}$
effectively equal to zero in the positive bias regime, the regime
of Fig.~\ref{fig2}.
%Note that
From Eqs.~(\ref{current}) and (\ref{e:cdagcevol}),
we see that Eq.~(\ref{Ist}) also equals to
the steady state current through the right junction as:
$I_{\rm st}=e\gamma_R \langle c^\dagger c \rangle_{\rm st}$.
In the case $\gamma_R=\gamma_L$,
the steady state $\langle c^\dagger c\rangle_{\rm st}$ shown in
Fig.~\ref{fig2}(a) and (d) at long times should thus equal respectively to
$I_{\rm st}/(e\gamma_L)$ shown in Fig.~\ref{fig2}(c) and (f) at long
times. This is indeed the case,
although the transient behaviors in these plots
differ considerably. We note that
the plot shown in Fig.~\ref{fig2}(c) and (f) is the current through
the left junction, normalized by $(e\gamma_L)$.
The values of $f_{1L}$, $f_{2L}$, and $f_{3L}$
depend on the strength of the applied bias voltage and are important in
understanding the stepwise behavior of the
stationary current as a function of the bias voltage.
We will now concentrate exclusively on the positive bias regime.

When the bias voltage is small, the current is zero. As the bias
voltage increases the first Fermi factor in the left lead
to become non zero is $f_{2L}$, with the other Fermi factors very
small or zero. In the case that $f_{2L}=1,\  f_{1L}=0,\ f_{3L}=0$,
the steady state current is
\begin{equation}
I^{(1)}_{\rm st}=e\gamma_L[\lambda^2\langle a^\dagger a\rangle_{\rm CT,
  st}-(1-\lambda^2)\langle c^\dagger c\rangle_{\rm st}]\ ,
\end{equation}
For low temperatures this is very small. Only if the phonon
temperature is large, so that the stationary mean phonon number is
significant, does this first current step become apparent
(see Fig.~\ref{fig4}). As the bias voltage is increased both $f_{2L}$ and
$f_{1L}$ become non zero. In the case where they are both unity, the
steady state current  is
\begin{equation}
I^{(2)}_{\rm st}=e\gamma_L[(1-\lambda^2)\langle
cc^\dagger\rangle_{\rm st}-\lambda^2\langle a^\dagger a\rangle_{\rm CT, st}]\ .
\end{equation}
The first term here is the same form as the bare tunneling case
except that the effective tunneling rate is reduced by
$(1-\lambda^2)$. This is not too surprising. If the island is moving,
on average it reduces the effective tunneling rate across the two
barriers.  Thus the first current step will be reduced below the value
of the bare (no phonon) rate.
At the region where bias voltage is large, all the Fermi factors are
unity and
\begin{equation}
I^{(3)}_{\rm st}=e\gamma_L\langle cc^\dagger\rangle_{st}\; ,
\end{equation}
which is the expected result for the bare tunneling case.

To explicitly evaluate the stationary current we need to solve for the
stationary mean electronic and phonon occupation numbers. We have done
this numerically and the results are shown in the figures
below. However the large bias case can be easily solved;
\begin{eqnarray}
\langle c^\dagger c \rangle_{\rm st} &=&
\frac{\gamma_L}{\gamma_L+\gamma_R} \; ,
\label{cdclg}\\
\langle a^\dagger a \rangle_{\rm st} &=&
\biggl(\lambda^2+\frac{\lambda^2(-\gamma_L+\gamma_R)}{\kappa}
\biggr)\biggl(\frac{\gamma_L}{\gamma_L+\gamma_R}\biggr) +
\frac{\gamma_L \lambda^2 + \kappa \bar{n}_p}{\kappa} \; ,
\label{adalg}\\
I_{\rm st} &=& e\frac{\gamma_L \gamma_R}{\gamma_L+\gamma_R}\; .
\label{Ilg}
\end{eqnarray}
This is the result for tunneling through a single quasi bound state
between two barriers \cite{Sun}.

The steady state current for larger values of $\lambda$ shows
two steps. As one can see from Fig.~\ref{fig2}(a),
the current vanishes until the
first Coulomb Blockade energy is overcome. The first step in the
stationary current is thus due to bare tunneling though the dot. The
second step represents single phonon mediated tunneling through
the dot. These results are consistent with the semiclassical
theory of Boese and Schoeller\cite{Boese} given that our expansion
to second order in $\lambda$ can only account for single phonon
events. The height of the step depends on $\lambda$ which is the
ratio of the coupling strength  between the electron and the
vibrational level $\chi$, and the oscillator energy $\hbar
\omega_o$.

Looking at Fig.~\ref{fig2}, the average electron number
approaches a steady state (e.g., a steady state value of $0.5$ at large
bias since we have set the
value of $\gamma_L$ to be equal to $\gamma_R$) while the average
phonon number, without external damping, behaves differently in
various regions (Fig.~\ref{fig2}(b)).
The average phonon number slowly reaches a steady
state value within the first step, while at the second step, the
mean phonon number grows linearly with time (Fig.~\ref{fig2}(b)).
The steady state at the first step is the previously
noted effect of transport induced damping.  These results are as
expected since from Eq.~(\ref{e:mastereqnsimple}), the term
consisting $\gamma_{L2}\lambda^2f_{2L}{\cal D}[ac^\dagger]\rho$
corresponds to a jump of one electron onto
the island with the simultaneous annihilation of a phonon,
while $\gamma_{L3}\lambda^2f_{3L}{\cal D}[a^\dagger c^\dagger]\rho$
corresponds to the jump of an electron onto the island along with the
simultaneous creation of a phonon. The dynamics can be understood by relating
the behavior of these terms to the rate
of average phonon number change in Eq.~(\ref{e:adagaevol}). At
the first step, when both $f_{1L}$ and $f_{2L}$ are both one, while
$f_{3L}$ is zero, the coefficient $B_2$ has negative value, and
therefore the mean phonon number could
reach a steady state under this transport induced damping.
In this regime, we find that
\begin{eqnarray}
\langle c^\dagger c\rangle_{\rm st}&=&(1-\lambda^2)/2\; ,
\\
\langle a^\dagger a\rangle_{\rm st}&\approx& 1/2\; ,
\label{ada2}
\end{eqnarray}
where we have set $\gamma_L=\gamma_R$ and used
Eq.~(\ref{CTadaTr}) in obtaining Eq.~(\ref{ada2}).
The corresponding effective temperature can be found using
\begin{equation}
  \label{eq:Teff}
  T_{\rm eff}=\frac{\hbar \omega_o}{k_B \ln [1+(1/\langle a^\dagger
  a\rangle_{\rm st})]} \; .
\end{equation}
When all the
Fermi factors for the left lead are unity,
the rate of growth for the mean phonon number
now depends on a constant $C_2$ in Eq.~(\ref{e:adagaevol})
and therefore the mean
phonon number will grow linearly with time.
However, the current will be still steady (Fig.~\ref{fig2}(f)).
This indicates that the steady state current
and mean electron
number in the dot [given by Eqs.~(\ref{Ilg}) and
(\ref{cdclg}) respectively] do not depend on the mean phonon number.
This is supported by the fact that the coefficient
$B_1$ in Eq.~(\ref{e:cdagcevol}) vanishes in this regime.
%This behaviour shows
%that voltage across the junction can act like a thermal bath for the
%oscillator.
When damping is included, the phonon number reaches a steady value of 0.35
[see Fig.~\ref{fig2}(e) and also Eq.~(\ref{adalg})].

\begin{figure}
   \begin{center}
%   \begin{tabular}{c}
   \includegraphics[height=7cm]{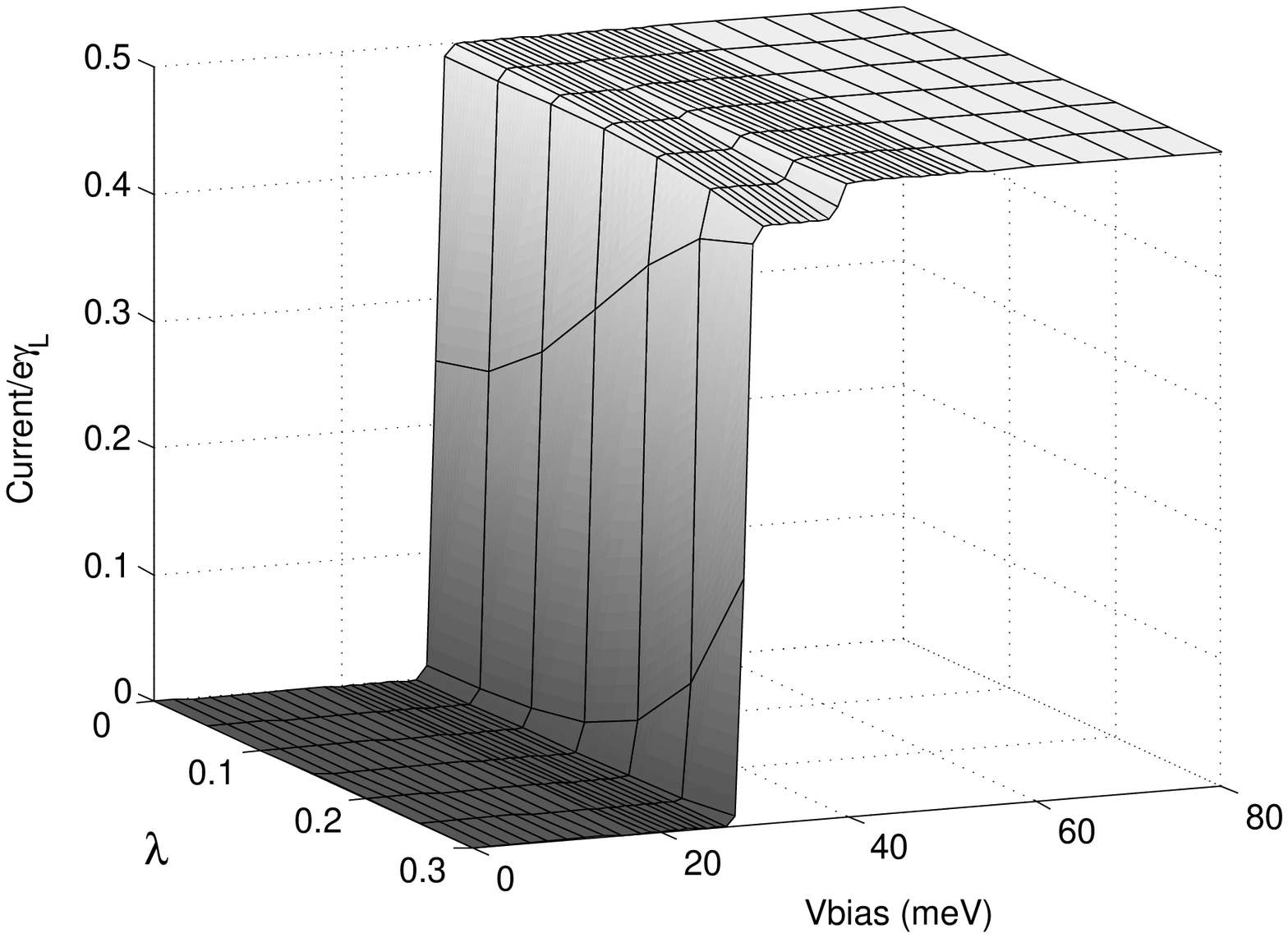}
%   \end{tabular}
   \end{center}
   \caption{Steady state current for different
       values of $\lambda$ with damping
     $\kappa = 0.3 \gamma_L$, the electronic and phonon temperatures are both $1.5$K.  \label{fig3}}
     \end{figure}

\begin{figure}
   \begin{center}
%   \begin{tabular}{c}
   \includegraphics[height=7cm]{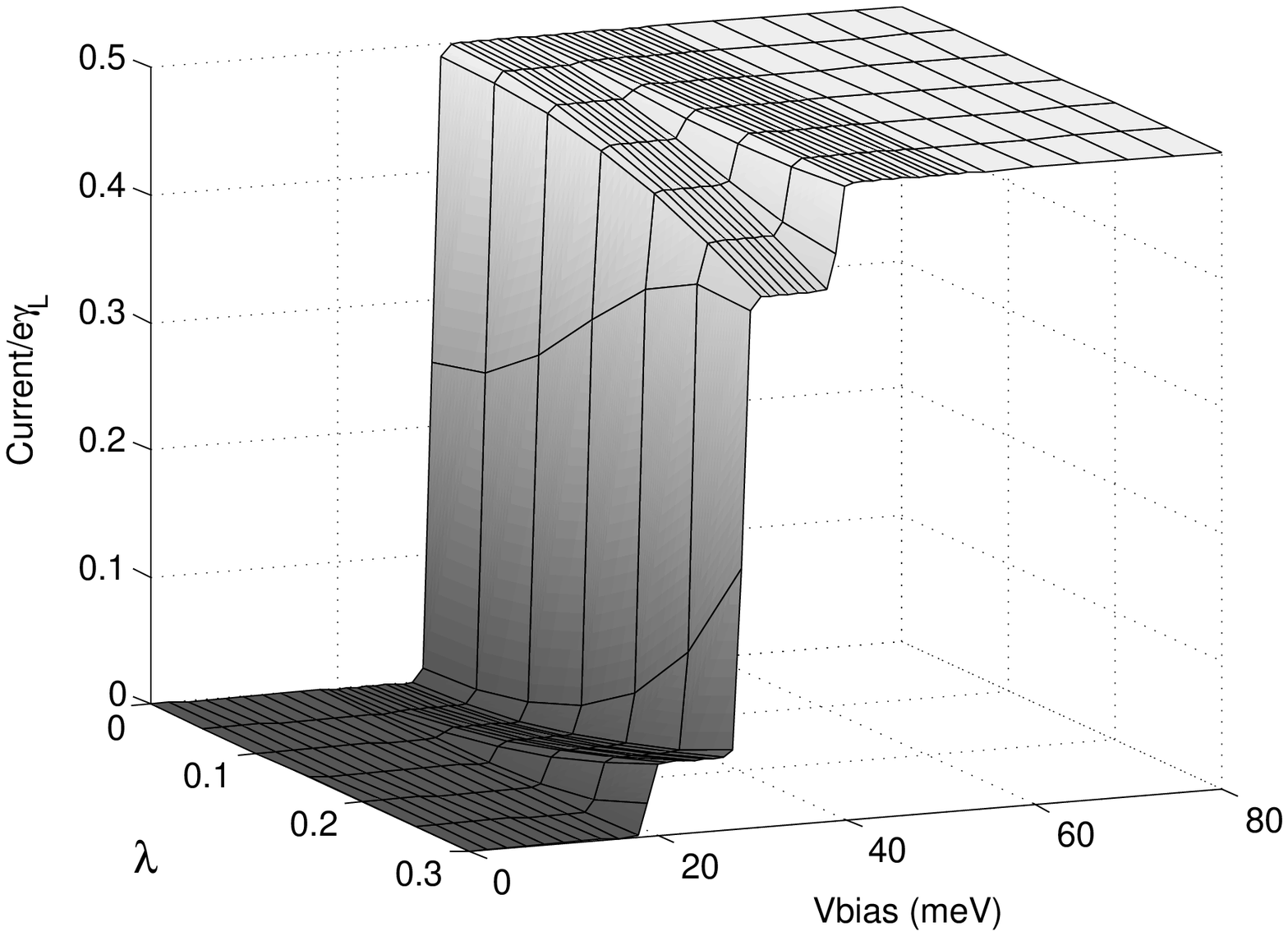}
%   \end{tabular}
   \end{center}
   \caption{Steady state current for different
       values of $\lambda$ and damping
     $\kappa = 0.3 \gamma_L$, the electronic temperature is $1.5$K, and the phonon temperature is $116$K which is chosen to be $2\hbar\omega_o/k_B$ in order
     to make manifest the step at a smaller bias voltage.  \label{fig4}}
     \end{figure}

In Fig.~\ref{fig3} we plot the steady state current versus bias
voltage for different values of $\lambda$. At the region of the
first phonon excitation level (when bias voltage is between 30 meV
and 40 meV), the steady state current drops by a factor
proportional to quadratic order of $\lambda$. Compared to the
current at large bias voltage,  the size of the drop is
\begin{eqnarray}
\Delta I_{\rm st} = \frac{\gamma_L \lambda^2 (2 \gamma_L \lambda^2 +
\kappa (2 \bar{n}_p +1))}{4 \gamma_L \lambda^2 - 2 \kappa
(\lambda^2 - 2)}.
\end{eqnarray}

We thus see that the effect of the oscillatory motion of the
island is two fold. Firstly the vibrational motion leads to an
effective reduction in the tunneling rate by an amount
$\lambda^2$ to lowest order in $\lambda$.   Secondly there is a
second step at higher bias voltage due to phonon mediated
tunneling. This is determined by the dependence of the Fermi
factors on the vibrational quantum of energy. One might have
expected another step at a smaller bias voltage. However this step
is very small unless there is a significant thermally excited mean
phonon number present in the steady state.  If we increase the
phonon temperature such that the energy is larger than the energy
quantum of the oscillator, $\hbar\omega_o$,
(for this example we choose $T=2\hbar\omega_o/k_B\approx 116K$ )
this step can be seen (Fig~\ref{fig4}).
Thus we see three steps as expected, corresponding to
the three bias voltages when the three Fermi factors switch from
zero.

\begin{figure}
   \begin{center}
%   \begin{tabular}{c}
   \includegraphics[height=7cm]{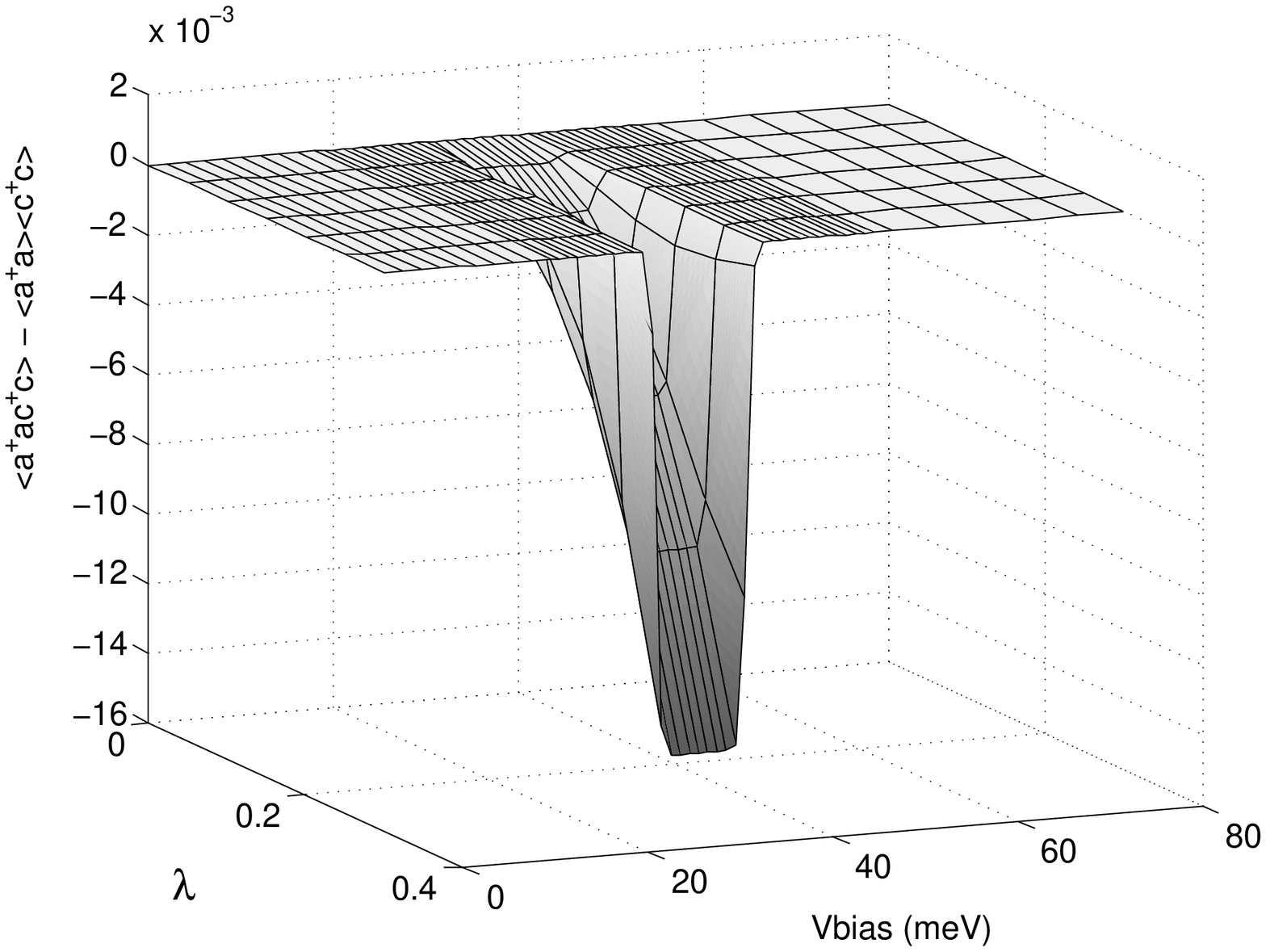}
%   \end{tabular}
   \end{center}
   \caption{Difference in the correlation function for different
       values of $\lambda$ with damping
     $\kappa = 0.3 \gamma_L$.  \label{fig5}}
%>>>> use \label inside caption to get Fig. number with \ref{}
   \end{figure}

\begin{figure}
   \begin{center}
%   \begin{tabular}{c}
   \includegraphics[height=7cm]{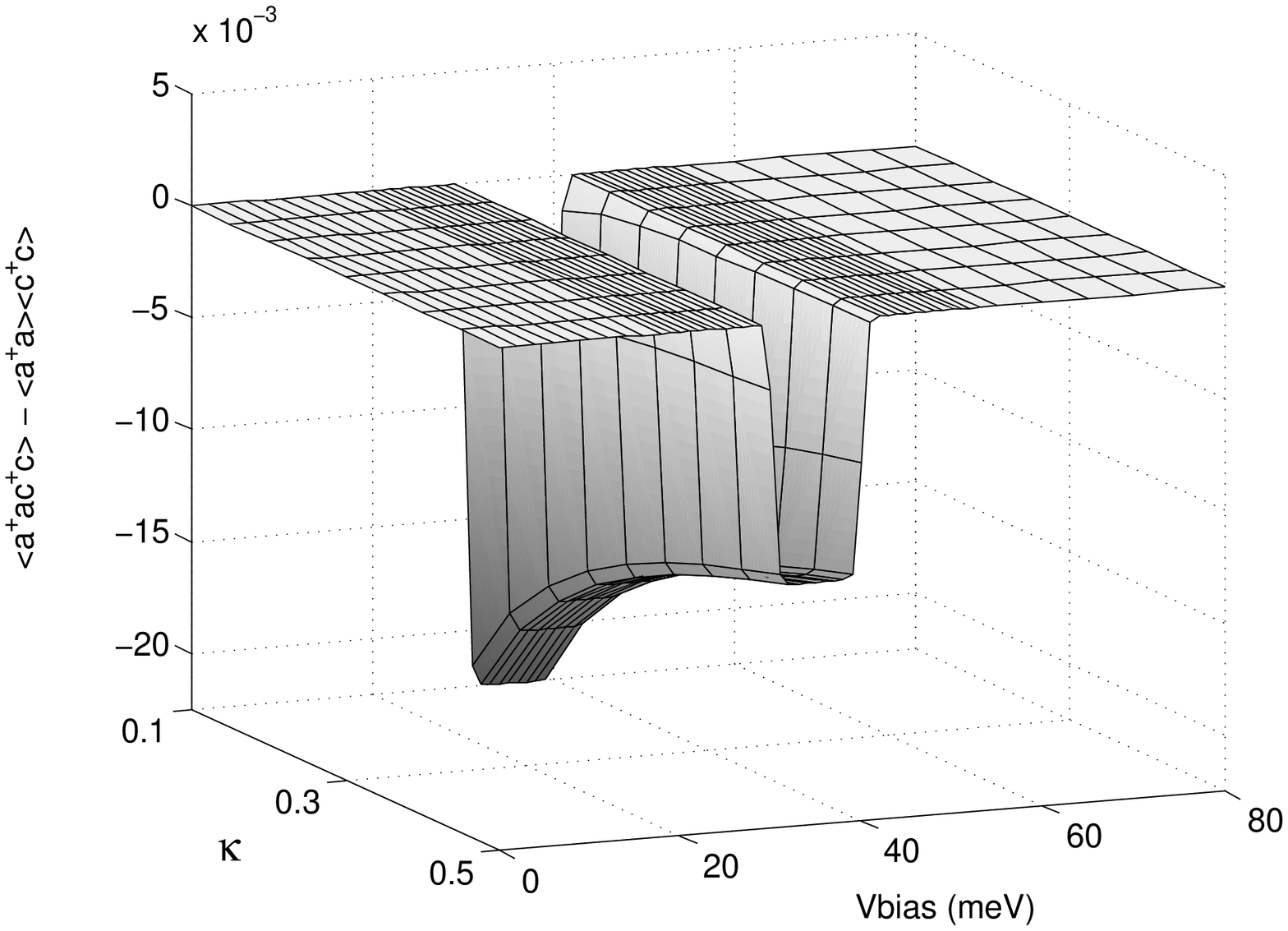}
%   \end{tabular}
   \end{center}
   \caption{Difference in the correlation function for different
     values of $\kappa$ with damping
     $\lambda = 0.3 \gamma_L$. The plot starts at a value $\kappa\neq
     0$.
\label{fig7}}
%>>>> use \label inside caption to get Fig. number with \ref{}
   \end{figure}

In order to explore the steady state correlation between phonon number
and electron number on the dot, we
can find the steady state directly by solving the
master equation from Eq.~(\ref{e:mastereqnsimple}).
In Fig.~\ref{fig5} we plot the correlation function
$\langle a^\dagger a c^\dagger c\rangle-\langle a^\dagger a\rangle\langle c^\dagger c\rangle$ as a function of $\lambda$ and the bias voltage.
The correlation is seen to be small except when a transition occurs
between two conductance states. This is not surprising, as at this
point one expects fluctuations in the charge on the dot, and
consequently the fluctuations of phonon number,  to be large. This
interpretation implies that damping of the oscillator should suppress
the correlation, as the response of the oscillator to fluctuations in
the dot occupation are suppressed. This is seen in Fig.~\ref{fig7}.

\section{Conclusions}
We have given a quantum description of a QEMS
comprising a single quantum dot harmonically bound between two
electrodes  based on a quantum
master equation for the density operator of the electronic and
vibrational degrees of freedom. The description thus incorporates the dynamics
of both diagonal (population) and off-diagonal (coherence) terms.
We found a special set of parameters for which
the equations of motion for the mean phonon number and the electron
number form a closed set.
From this we have been able to reproduce  the central qualitative
features of the current vs.~bias voltage curve obtained
experimentally by Park et al.\cite{Park} and also of the semiclassical
phenomenological
theory by Boese and Schoeller \cite{Boese}.   We also calculate the
correlation
function between phonon and electron number in the steady state
and find that it is only significant at the steps of the the steady
state conductance.
The results reported in this paper do not probe the full power of the
master equation approach as
the model does not couple the diagonal and off-diagonal elements
of the density matrix. This can arise when the vibrational motion of
the dot is subject to a
conservative driving force, in addition to the stochastic driving
that arises when electrons tunnel on and off the dot in a static
electric field.
The full quantum treatment will enable us to include coherent effects which
are likely to arise when a spin doped quantum dot is used in a
static or RF external magnetic field.

%%%%%%%%%%%%%%%%%%%%%%%%%%%%%%%%%%%%%%%%%%%%%%%%%%%%%%%%%%%%%
%%%%%Sometimes it is necessary to precede the double slash
%%%%%by \verb|\protect| to get the desired result,
%%%%%for example, in article titles.

%%-----------------------------------------------------------

\section*{Acknowledgments}
G.J.M and D.W.U acknowledge the support of the Australian Research
Council Federation Fellowship Grant. H.S.G would like to
acknowledge support from Hewlett-Packard. We gratefully
acknowledge discussions with Jason Twamley, supported by EC IST
FET project IST-2001-37150 QIPDF-ROSES.
%%-----------------------------------------------------------
%%%%%%%%%%%%%%%%%%%%%%%%%%%%%%%%%%%%%%%%%%%%%%%%%%%%
\appendix    %>>>> this command starts appendixes
%%%%%%%%%%%%%%%%%%%%%%%%%%%%%%%%%%%%%%%%%%%%%%%%%%%%

\bibliographystyle{prsty}

\begin{thebibliography}{99}
\bibitem{roukes}M.~Roukes, Physics World {\bf 14}, 25, February (2001).
\bibitem{knobel-clleland}R.G. Knobel and A.N. Cleland, Nature {\bf
    424}, 291 (2003).
\bibitem{Schwab04}M.~D. LaHaye, O.~Buu, B.~Camarota, and K.~C.~Schwab,
    Science {\bf 304}, 74 (2004).
\bibitem{spin-detection}Z.~Zhang, M.~L.~Roukes, and P.~C.~Hammel,
  Journ. App. Phys. {\bf 80}, 6931 (1996); H.~J. Mamin, R.~Budakian,
  B.~W.~Chui and D.~Rugar, Phys. Rev. Lett. 91, 207604 (2003).
\bibitem{Brun03}
T.~A.~Brun and H.-S.~Goan, Phys. Rev. A {\bf 68}, 032301 (2003);
G.~P.~Berman, F.~Borgonovi, H.-S.~Goan, S.~A.~Gurvitz and
V.~I.~Tsifrinovich, Phys. Rev. B {\bf 67}, 094425 (2003).


\bibitem{quantum-effects}A.~D.~Armour, M.~P.~Blencowe, and
  K.~C.~Schwab,  Phys. Rev. Lett. {\bf 88}, 148301 (2002).
\bibitem{Huang}X.~M.~H.~Huang, C.~A.~Zorman, M.~Mehregany, and
  M.~Roukes,  Nature {\bf 421}, 496 (2003).
\bibitem{Park}H.~Park, J.~Park, A.~K.~Lim, E.~H.~Anderson,
  A.~P.~Allvisator, and P.~L. McEuen, Nature {\bf 407}, 57, (2000).
\bibitem{Mozyrsky-Martin}D.~Mozyrsky and I.~Martin, Phys. Rev.
Lett. {\bf 89}, 018301 (2002).
\bibitem{Smirnov}A.~Yu.~Smirnov, L.~G.~Mourokh, and N.~J.~M.~Horing,
Phys. Rev. B {\bf 67}, 115312(2003)
\bibitem{Mozyrsky-Martin-Hastings}D.~Mozyrsky, I.~Martin and
  M.~B.~Hastings, Phys. Rev. Lett. {\bf 92}, 018303 (2004).
\bibitem{Armour} A.~D.~Armour, M.~P.~Blencowe, and Y.~Zhang, Phys. Rev. B
  {\bf 69}, 125313 (2004).
%\bibitem{Milburn}G.J. Milburn and C.A.Holmes,
%Phys. Rev. Lett. {\bf 56}, 2237 (1986).
\bibitem{Boese}D.~Boese and H.~Schoeller, Europhys. Lett. {\bf 54}, 668 (2001).
\bibitem{Sun}H.~B.~Sun and G.~J.~Milburn,  Phys. Rev. B. {\bf 59},
  10748 (1999); G.~J.~Milburn, Aust. J. Phys. {\bf 53}, 477 (2000).
\bibitem{Aji}V.~Aji, J.~E.~Moore and C.~M.~Varma,
 {\em Electronic-vibrational coupling in single-molecule devices}, cond-mat/0302222 (2003).
\bibitem{Gardiner-Zoller}C.~W. Gardiner and P.~Zoller, {\em Quantum Noise}, 2nd edition, (Springer-Verlag, Berlin,  2000)
\bibitem{shuttle-rev}R.~I.~Shekhter, Yu.~Galperin, L.~Y.~Gorelik,
  A.~Isacsson and M.~Jonson, J. Phys.: Condens. Matter {\bf 15}  R441
  (2003).
\bibitem{mahan}G.~Mahan, {\em Many-Particle Physics}, (Plenum Press, 1990).



\end{thebibliography}

\end{document}